\documentclass{jaa}
\usepackage{txfonts}
\usepackage{graphicx}
\usepackage{color}

\usepackage[authoryear]{natbib}

\usepackage{txfonts}
\usepackage[T1]{fontenc}

\DeclareRobustCommand{\VAN}[3]{#2}
\let\VANthebibliography\thebibliography
\def\thebibliography{\DeclareRobustCommand{\VAN}[3]{##3}\VANthebibliography}

\usepackage{graphicx}

\usepackage{ae,aecompl,bm}
\usepackage{xcolor}
\usepackage{subcaption}

\usepackage{tabularx}
\usepackage{natbib}

\date{}
\def\be{\begin{equation}}
\def\ee{\end{equation}}

\begin{document}\sloppy

\title{Conversion of Emitted Axionic Dark Matter to Photons for Non-Rotating Magnetized Neutron Stars}

\author{Shubham Yadav\textsuperscript{1}\thanks{e-mail: shubhamphy28@gmail.com}, M. Mishra\textsuperscript{1}\thanks{e-mail: madhukar@pilani.bits-pilani.ac.in},Tapomoy Guha Sarkar\textsuperscript{1}\thanks{e-mail: tapomoy1@gmail.com}}
\affilOne{\textsuperscript{1}Department of Physics, Birla Institute of Technology and Science, Pilani, Rajasthan, India\\}

\onecolumn{
\maketitle

\begin{abstract} 

We attempt to find the impact of a modified Tolman–Oppenheimer–Volkoff (TOV) system of equations  on the luminosities of direct photons, neutrinos \& axions for a particular axion mass in the presence of a magnetic field. We employ two different equation of states (EoSs) namely APR \& FPS to generate the  profiles of mass and pressure for spherically symmetric and non-rotating Neutron stars (NSs). We then compute the axions and neutrino emission rates by employing the Cooper-pair-breaking and formation process (PBF) in the core using the NSCool code. We also examine the possibility of axion-to-photon conversion in the magnetosphere of NSs. Furthermore, we investigate the impact of the magnetic field on the actual observables, such as the energy spectrum of axions and axion-converted photon flux for three different NSs. Our comparative study indicates that axions energy spectrum and axion-converted photon flux changes significantly due to an intense magnetic field.

\end{abstract}

}

\doinum{12.3456/s78910-011-012-3}
\artcitid{\#\#\#\#}
\volnum{000}
\year{0000}
\pgrange{1--}
\setcounter{page}{1}
\lp{1}

\section{Introduction}
\label{intro}
Various astrophysical observations indicates that a significant fraction ($\sim 30 \%$) of the universe's matter-energy budget is in the form of dark matter (DM). There has been a tremendous effort to understand this mysterious matter in terms of particles and in the framework of particle physics. The QCD axion is a promising candidate for such DM \cite{PhysRevD.100.083005,PhysRevC.98.035802,Dessert_2022,PhysRevLett.123.021801,paul2019neutron}. These hypothetical particles are postulated to explain the CP conservation in QCD, commonly called a strong CP problem. The axion field is introduced through a derivative coupling to a fermion field $\psi_{f}$ with an interaction Lagrangian ${\cal{L}}_{int} = ( C_f/2\,f_a) \bar{\psi}_f \gamma^{\mu} \gamma_{5} \psi_f  \partial_{\mu} a$, where $f_{a}$ is the decay constant of axion. This term allows for the conversion of a fermion to an axion. Two possible axion models exist in the literature: the Kim-Shifman-Weinstein-Zakharov - KSVZ (hadronic) and the Dean-Fischler-Srednitsky-Zhitnitsky - DFSZ model, depending on whether the axions couple only with hadrons or leptons ~\cite{Leinson_2019}. Astrophysical and cosmological constraints put bounds on axion rest mass as: $m_{a} \sim 10^{-6}$ eV to $10^{-2}$ eV.
 
Several theoretical and experimental attempts have been carried out so far to find the properties of the axions and to explore their detection possibilities~\cite{Kaminker_2006,umeda1998axion,adams2022axion,PhysRevD.97.123006}. The Axionic DM experiment (ADMX) located at the University of Washington, US is one such experiment and will cover much of the axions parameter space. According to the ADMX experiment, if these very light particles exist, then it could be possible that they could decay into a pair of light particles, thus making them difficult to detect. Also the various collider experiments~\cite{PhysRevLett.118.171801,GONCALVES2020135981,ZHU2022115961} continues to search for these weakly interacting particles so-called axions. 

 At the end stage of the star's life, massive stars undergo a violent transition to produce compact objects such as NSs, White Dwarfs and Black holes. NSs provide an excellent laboratory for constraining the properties of light and weakly interacting QCD axions. Axions may be produced in the core \& (crust) of a NSs through Cooper pair formation/breaking mechanism~\cite{Sedrakian_2016,Keller_2013,Buschmann:2019pfp} and bremsstrahlung in nucleon (\& electron) scattering processes ~\cite{Sedrakian_2016,Buschmann:2019pfp,paul2019neutron}.
 
The axions emitted from the core/crust may resonantly convert into X-ray photons due to a strong magnetic field inside the magnetospheres of the NSs ~\cite{Buschmann:2019pfp}. Axions in extended NSs magnetospheres can couple to virtual photons and produce real photons due to the Primakoff effect \cite{Pshirkov_2009}.
Recent observations~\cite{DEXHEIMER2017487,Dexheimer:2017fhy,Lopes_2015,2019PhRvC..99e5811C} suggests about a specific group of NSs, called Magnificent Seven (magnetars). Magnetars are strongly magnetized NSs that exhibit a wide array of X-ray activity, with extraordinarily strong magnetic field intensities of up to $10^{18}$ Gauss. In the literature, many authors have proposed that the internal structure and cooling properties are affected by the distribution of a strong magnetic field in the interior of NSs. 
The emission properties of highly magnetized stars get modified due to the change in the stellar structure equations (TOV equations) and the significant change in the conductive and convective processes in the heat blanketing layer of the NSs. It is found that the NS cooling shall be affected by the strong magnetic field, which will further affect the emission of axions and their  subsequent conversion to photons. Furthermore, the equation of state (EoS) of the NSs shall also have an imprint on the emission properties of various observables as a result of NS cooling. 

Safdi et al.~\cite{PhysRevD.99.123021} examined the possibility of axion-photon conversion in NSs with astrophysical galaxy clusters. They have described the NSs distribution by including population synthesis and evolution models. Hook et al.~\cite{PhysRevLett.121.241102} reported observations of radio signals from axionic DM. They have solved axions-photons mixing equations in magnetized plasma. The axions or ALP-to-photon conversion probability in the magnetosphere is calculated in detail by Fortin et al.~\cite{fortin2018constraining} by reshaping the coupled differential equations of ALP-photon propagation. In the process, they have created an analytic formalism to perform comparable computations in broader aspects of n-state oscillation systems. G. Raffelt et al. created a framework that may be used to examine how a photons (axions, gravitons) beam changes in the presence of outside influences such as strong magnetic field and gravity. They have used their findings to discuss the detection of axions by measuring the magnetically induced birefringence of the vacuum. 

Francesco D'Eramo presents noteworthy advancements in context to axions from the early universe by calculating the axion emission rate for various scenarios, and utilizing these findings to forecast the number of axions produced. They also revised cosmic limits on the mass range of the QCD axion.

In the current work, we have investigated the impact of magnetic field on the axion energy spectrum and axion-converted photon flux in the magnetosphere of the non-rotating NSs for mass M $\sim 1.4M_{\odot}$. We have used a radially varying intense magnetic field in the modified TOV system of equations to generate the profiles for mass, pressure,\& baryon density. Here, we have first determined the impact of the magnetic field on the luminosities of emitted particles i.e. direct photons, neutrinos, and axions emitted from the core of the NSs via the PBF process using the NSCool code~\cite {2016ascl.soft09009P}. We then calculated its conversion into photons due to axion-photon coupling in the presence of the high magnetic field in the magnetospheres of NSs. We employ two EoSs, namely APR \cite{PhysRevC.58.1804,Gusakov_2005}, and FPS  ~\cite{Flowers:1976ux}. In addition, we also have calculated axion energy spectrum and the axion-converted photon flux. Finally, we conclude that the magnetic field affects the axion energy spectrum and the axion-converted photon flux due to the change in the mass and pressure profiles of the star (the effect of magnetic field incorporated via the TOV system of equations).\\

\noindent The paper is structured as follows: In Section~\ref{intro} we present the Introduction. Section~\ref{form} elaborates on the modified TOV equations, Neutron star cooling, Axion physics and, Axion \& Neutrino emission rates in the core of NSs. In Section~\ref{result}, We discussed how including a magnetic field affects the luminosities (neutrinos, axions, and direct photons) for a particular axion mass. We also discuss the obtained numerical results on the impact of intense magnetic fields for various observables and macroscopic NS properties.  Further, we describe the energy spectrum of axions and discuss axion-converted photon flux. Finally, in Section \ref{conc}, we summarize our results and conclude the work. 

\section{Formalism}
\label{form}
\subsection{Modified TOV Equations for non-rotating NSs}
The stellar equations get modified in the presence of a magnetic field as a these strong magnetic field contributes to the energy-momentum tensor and thereby changes the TOV system of equations~\cite {2019PhRvC..99e5811C}, energy balance, and heat transport equations. The modified TOV equations in the presence of a magnetic field are given by: 
\begin{equation}
\frac{dm}{dr}=4\pi r^{2}\left ( \epsilon +\frac{B^{2}}{2\mu _{0}c^2} \right )
\end{equation}
\begin{equation}
\frac{d\phi }{dr}=\frac{G\left ( m(r)+  4 \pi r ^3  P/c^{2} \right )}{r\left ( r\,c^2-2G\,m(r)\right )}
\end{equation}
\begin{equation}
\frac{dP}{dr}=-c^{2} \left ( \epsilon +\frac{B^2}{2\mu _{0}c^2}+\frac{P}{c^{2}} \right )\left ( \frac{d\phi }{dr}-{\mathcal L}\left ( r \right ) \right ),
\end{equation}
where $\mathcal L(r)$  denotes a  Lorentz force contribution~\cite {2019PhRvC..99e5811C}.

\noindent Under extreme conditions of very high density, apart from the magnetic field distribution in a NSs~\cite{dexheimer2012hybrid,Reddy:2021rln,1995A&A...301..757B}, the EoS of matter also plays a crucial role in determining the internal structure and cooling properties of NSs~\cite {PhysRevD.37.2042,PAGE2006497,refId0,YAKOVLEV2004523,10.1093/mnras/sty776,10.1093/mnras/stx366}.  
  
\noindent The detailed study for composition of NSs matter is strongly analysed by the EoS of the matter. Several EoSs have been reported with and without considering finite baryon chemical potential~\cite{DEXHEIMER2017487,10.1093/mnras/stu2706,Lattimer_2001,Schneider_2019}. 

By considering the azimuthal symmetry the magnitude of the magnetic field $B = ({\vec B} .{\vec B})^{1/2}$ is expressed as a sum:
\begin{equation}
B(r, \theta) \approx \sum_{l=0}^{l_{max}} B_{l} (r)  Y_l^0 (\theta). 
\end{equation}
We shall ignore the anisotropy of the magnetic field as a first approximation in our work. We note  that the vector magnetic field $B$ doesnot have any monopole component. Monopole term is related to the norm of the magnetic field. In the presence of strong magnetic fields the stability analysis of NSs requires the presence of both poloidal \& toroidal fields~\cite{10.1111/j.1365-2966.2008.14034.x}. In principle, it has also been reported that the effects of toroidal fields are much larger than poloidal fields for such a realistic stars~\cite{10.1111/j.1365-2966.2008.14034.x,2006A&A...450.1077B}.
Here, a simplified pure radial profile is used for the monopolar term of $B$ (we emphasise that we are not including magnetic monopoles here) as a polynomial fit function~\cite{2021EPJC...81..698P,Psaltis_2014} which is given by:
\begin{equation}
B_0(r) = B_c\left [ 1 - 1.6 \left(\frac{r}{\bar r} \right )^2  -  \left(\frac{r}{\bar r} \right )^4 + 4.2  \left(\frac{r}{\bar r} \right )^6 -2.4  \left(\frac{r}{\bar r} \right )^8 \right],
\end{equation}
where $\bar r$ is the star's mean radius. Finding a suitable radius of a distorted star due to a magnetic field is a bit complicated. Therefore, the mean radius $\bar{r}$ is directly related to the star's surface. We have adopted a fiducial value of $B_c = 10^{18}G$ in our work. We have used this as a universal magnetic field profile and neglected all the possible variations that may arise from different EoSs.

An effective Lorentz force term in the modified TOV equations~\cite {2019PhRvC..99e5811C,10.1093/mnras/sty776}is given by:
\begin{equation}
\frac{{\mathcal L}\left ( r \right )}{ 10^{-41}} = B_{c}^2\left[-3.8 \left(\frac{r}{\bar r} \right ) + 8.1 \left(\frac{r}{\bar r} \right )^3  -1.6   \left(\frac{r}{\bar r} \right )^5 -2.3   \left(\frac{r}{\bar r} \right )^7 \right ]
\end{equation}

Here, we employ two different EoSs namely, APR and FPS to study the thermodynamic behavior of the   NSs interiors. In Akmal-Pandharipande-Ravenhall (APR)~\cite{Gusakov_2005,PhysRevC.58.1804}, EoS interaction potential is parameterized using baryon density and isospin asymmetry~\cite{Haensel_2002,Schneider_2019}. This EoS presents the transition between the low-density phase (LDP) to the high-density phase (HDP). It is also observed that phase transition in this EoS boosts up the shrink rate of the star. The FPS EoS gives a unified description of the inner crust and the liquid core. In this model, the liquid-crust core transition takes place at higher density ~$\rho_ {edge}$ = $1.6 \times 10^{14}$ gm/cm$^{3}$ and it is preceded by a series of phase transitions between various nuclear shapes. These transitions implies a gradual stiffening of matter.

\subsection{Neutron star cooling}
\label{nsc}
The extensive study of superdense matter is carried out by solving the heat transport and energy balance equations in full General Relativity. The modified TOV equations in the presence of a magnetic field are initially solved for a given EoS to generate pressure, density, and mass profiles. The cooling simulations relevant for this work is carried out by using the NSCool code \cite{2016ascl.soft09009P}.
Assuming the star interiors as an isothermal and spherically symmetric, and also by considering energy conservation equation, one can have a adequate understanding for the cooling of compact stars~\cite{Beznogov_2023,Yakovlev_2005, PhysRevD.37.2042, PAGE2006497, YAKOVLEV2004523, PhysRevLett.106.081101,PhysRevLett.120.182701,Heinke_2009}. The energy conservation equation of star follows the Newtonian formulation reads as~\cite{Buschmann:2021juv}:
\begin{equation}
C_{v}\frac{\mathrm{dT_b^{\infty}} }{\mathrm{d} t}=-L_{\nu }^\infty-L_{a}^\infty-L_{\gamma }^\infty(T_{s}^{\infty})+H,
\end{equation}
where
$L_{\nu }^\infty$ is the neutrino luminosity,  
$C_{v}$ total specific heat,
$L_{\gamma }^\infty$ is surface photon luminosity,
$L_{a}^\infty$ is the axion luminosity, 
 $T_{s}$ is the surface temperature 
$H$ includes all possible heating mechanisms,
and  $T_b^{\infty}$ internal temperature.
$L_{\gamma }^\infty$= $4\pi\sigma R^{2}(T_{s}^{\infty})^4$, with $\sigma$ as Stefan-Boltzmann's constant and $R$ is the radius of the star.  $T_{s}^{\infty}$ = $T_{s}\sqrt{1-2GM/c^{2}R}$, Here infinity superscript indicates that the external observer is at infinity and measures these quantities on the redshifted scale. Typically, $T_{s}^{\infty}/T_{s}$ $~\sim$ $0.7$. Our analysis assumes superfluidity in the core and $H=0$ in the entire NSs cooling.

\subsection{ Neutrino-Axion emission inside NS core}
\label{aer}
\subsubsection{Cooper Pair Breaking and Formation process (PBF)}
\label{cooper}
In the current work, we have only considered $^1S_0$ pairing for both neutrons and protons. This pairing usually occurs at densities which corresponds to the core of NSs. Cooper pairs are expected to form and break when the n-n $\&$ p-p superfluids are in thermal equilibrium with the broken pairs excitations. This happens only when temperature T$\ll$ $T_{c}$, where $T_{c}$ is defined as the superfluid critical temperature. The formation of these Cooper pairs liberates an energy carried by a $\nu\bar{\nu}$ pair.

\begin{align} 
X+X\to \left [ XX \right ] + \nu +\overline{\nu}\nonumber\\
\end{align}
Here, X can be neutrons or protons. The formation of these pairs releases an energy which is  carried out by the $\nu\bar{\nu}$ pair. 
\subsubsection*{Neutrino emission rate} 
\label{ner}
The NSs cools by emitting neutrinos through Cooper pair-breaking processes (PBF) with neutrons \& protons superfluid inside the core~\cite{Buschmann:2019pfp,Sedrakian_2016,Keller_2013,Page:2005fq,Page_2009}.This usually happens when the temperature~\cite{Geppert_2006,Kolomeitsev_2008,PhysRevLett.66.2701} lies below the critical temperature $T_{c}=10^{9}$ K. For neutron/proton $^1S_{0}$-wave paired superfluid, the emissivity~\cite{2001PhR...354....1Y,Keller_2013,Leinson_2000} of neutrino is given by:

\begin{equation}
\epsilon_{\nu }^s = \frac{5\,G_{F}^2}{14\pi^{3}}v_{N}\left ( 0 \right )v_{F}\left ( N \right )^2T^7I_{\nu}^s.
\end{equation}
Here the integral $I_{\nu}^s$ is;

\begin{equation}
I_{\nu}^s=z_{N}^7\left ( \int_{1}^{\infty }\frac{y^5}{\sqrt{y^{2}-1}} \left [ f_{F}\left ( z_{N}y \right ) \right ]^2 dy\right ),
\end{equation}
where $\epsilon_{\nu}^s$ is neutrino emissivity and $G_{F}$ is Fermi's coupling constant $=1.166\times 10^{-5} \mbox{GeV}^{-2}$. $z=\Delta(T)/T$ with $\Delta(T)=3.06\,T_c\,\sqrt{(1-\frac{T}{T_c})}$. Here $T_c $ is the critical temperature for neutron/proton superfluid. 
The corresponding neutrino emissivity due to the P-wave paired neutron superfluid is adapted from ref.~\cite{Page:2005fq}.\\

\subsubsection*{Axion emission rate}
In this section, we briefly describe the axion emission rate ~\cite{PhysRevLett.123.061104,PhysRevLett.53.1198} from PBF process inside NSs core~\cite{Sedrakian_2016}. In order to calculate the production rates within the NSs core, prerequisites are the temperature profiles in the core, the metric, the critical temperature profiles, neutron and proton Fermi's momenta profile (which inturn depends on the so-called equation of state). We have incorporated axion emission in the NSCool code to determine the luminosities of direct photons, neutrinos \& axions. 
Both spin-$0$ S-wave and spin $1$ P-wave nucleon superfluids could be possible inside the NS core. The axion emission rate~\cite{Buschmann:2019pfp,Sedrakian_2016} due to the neutron S-wave pairing from the Cooper pair-breaking formation (PBF)~\cite{Keller_2013} is given by:

\begin{equation}
\epsilon_{ax}^s= \frac{8}{3\pi f_{a}^{2}}\,v_{n}(0)\,v_F(n)^{2}\,T^{5}\,I_{ax}^s.
\end{equation}
The integral $I_{ax}^s$ is expressed as:

\begin{equation}
I_{ax}^s= z^5_{n}\left ( \int_{1}^{\infty} \frac{y^3}{\sqrt{y^2-1}}\left [ f_{F}\left ( z_{n}y \right ) \right ]^{2} dy\right),
\end{equation}
where
$\epsilon_{a}^s$ is axion emissivity,
$f_{a}$ is axion decay constant,
$v_{n}(0)$ density of state at Fermi surface and
$v_F(n)$ is the Fermi velocity of the neutron.

\begin{eqnarray}
v_{n}\left ( 0 \right )=\frac{m_{n}\,p_{F}\left ( n \right )} {\pi^2} \\
z=\frac{\Delta(T)}{T}, \\
f_{F}\left ( x \right )= \left [ e^x+1 \right ]^{-1},
\end{eqnarray}
with $x=\frac{\omega}{2T}$, where $\omega$ is the axion energy. \\

Finally, the ratio between axion and neutrino emissision rates is given as: 
\begin{equation}
{\epsilon _{a}^s} = \left ( \frac{59.2}{f_{a}^2G_{F}^2[\Delta\left ( T \right)]^2}r(z) \right )\epsilon_{\nu}^s,
\end{equation}
where, 
\begin{align}
\Delta(T)\simeq 3.06\,T_{cn}\sqrt{1-\frac{T}{T_{cn}}}.
\end{align}
The numerical values of the $r(z)=z^2\,I_{ax}^s/I_{\nu}^s$ associated with axion and neutrino emissivity integrals for the different values of $z$ are always less than or equal to unity.

\begin{equation}
f_{a}> 5.92\times 10^9 GeV\left [ \frac{0.1 MeV}{\Delta\left (T\right)} \right ].
\end{equation}
Here $f_{a}$ is a decay constant of axion~\cite{Buschmann:2021juv}. The relation between mass $m_{a}$ and decay constant $f_{a}$ is given by the equation;~\cite{Sedrakian_2016,Leinson_2019,PhysRevD.34.843,PhysRevD.37.1237}:

\begin{equation}
m_{a}= 0.60\; \text{eV}\times \frac{10^{7}\text{GeV}}{f_{a}}.
\end{equation}
Here, we have assumed the equal emissivity of axions for proton S-wave pairing and neutron S-wave pairing~\cite{Sedrakian_2016,1999A&A...343..650Y,1999A&A...345L..14K}. Also we have not considered any P-wave pairing in the work.\\

\subsection{Axion-Photon conversion}
\label{nnbrem}
\subsubsection{Energy Spectrum of Axions}
\label{cooper}
The energy spectrum of axions emitted due to the s-wave pairing is given by \cite{Buschmann:2019pfp}
\begin{equation}
{J_{ax,PBF}^{s}}= \frac {N_{ax,PBF}^{s}}{2\Delta T}\frac{\left ( \frac{\omega }{2\Delta T} \right )^3}{\sqrt{\left ( \frac{\omega }{2\Delta T} \right )^2-1}}\left [ f_{F}(\frac{\omega }{2T}) \right ]^2,
\end{equation}
where ${N_{ax,PBF}^{s}}$ = $\varepsilon_{ax}^{s}\times z_{n}^{5}/I_{ax}^{s}$ is the normalization constant derived from the expression $\int_{2\Delta T}^{\infty }J_{ax}^{s}d\omega =\epsilon_{ax}^{s}$ and $2y\Delta T$ is axion energy.
\subsubsection{Conversion Probability}
The axions, produced in the cores of NSs, are converted into X-rays in the NSs magnetospheres. The axions couple to photons given by an operator \cite{Buschmann:2019pfp}:
\begin{equation}
\mathcal L= -\frac{1}{4}g_{a\gamma \gamma }aF\widetilde{F}.
\end{equation}
Here, F \& $\widetilde{F}$ is the electromagnetic field strength tensor and corresponding dual field, respectively,$g_{a\gamma \gamma }$ is coupling constant.
Here, $g_{a\gamma \gamma }$ is given by:
\begin{equation}
g_{a\gamma \gamma }=\frac{C_{\gamma}\alpha}{2\pi f_{a}},
\end{equation}
where, $C_{\gamma}$ is dimensionless coupling constant and $\alpha=1/137$.
In the NSs magnetosphere, this operator can rotate an initial axion state to an electromagnetic wave which is polarised along the same direction as the external magnetic field. The approximate relation for conversion probability is given by:
\begin{equation}
\begin{split}
P_{a\to \gamma }\approx 1.5\times10^{-4}\left ( \frac{g_{a\gamma \gamma }}{10^{-11}GeV^{-1}} \right )^2\left ( \frac{1 keV}{\omega } \right )^{0.8} 
\times \left ( \frac{B_{0}}{10^{13}G} \right )^{0.4}\left ( \frac{R_{NS}}{10 km} \right )^{1.2}sin^{0.4}\theta
\end{split}
 \end{equation}
where, axions frequency $\omega \sim $ keV and $R_{NS}$ $\sim $10 km. 

Finally, we have obtained the photon flux by multiplying the axion energy spectrum by the probability of axion-photon conversion for different axion energies.  

\section{Results and Discussions}
\label{result}
The imprint of magnetic field on the energy spectrum of axions and the axions converted photon flux is studied by assuming the emission of various observables (axions, neutrinos and direct photons) from the core of the NSs. The present simulations are carried out by using the NSCool code~\cite{2016ascl.soft09009P}.\\

\begin{figure*}[htb!]
\centering
\includegraphics[width=7.75cm]{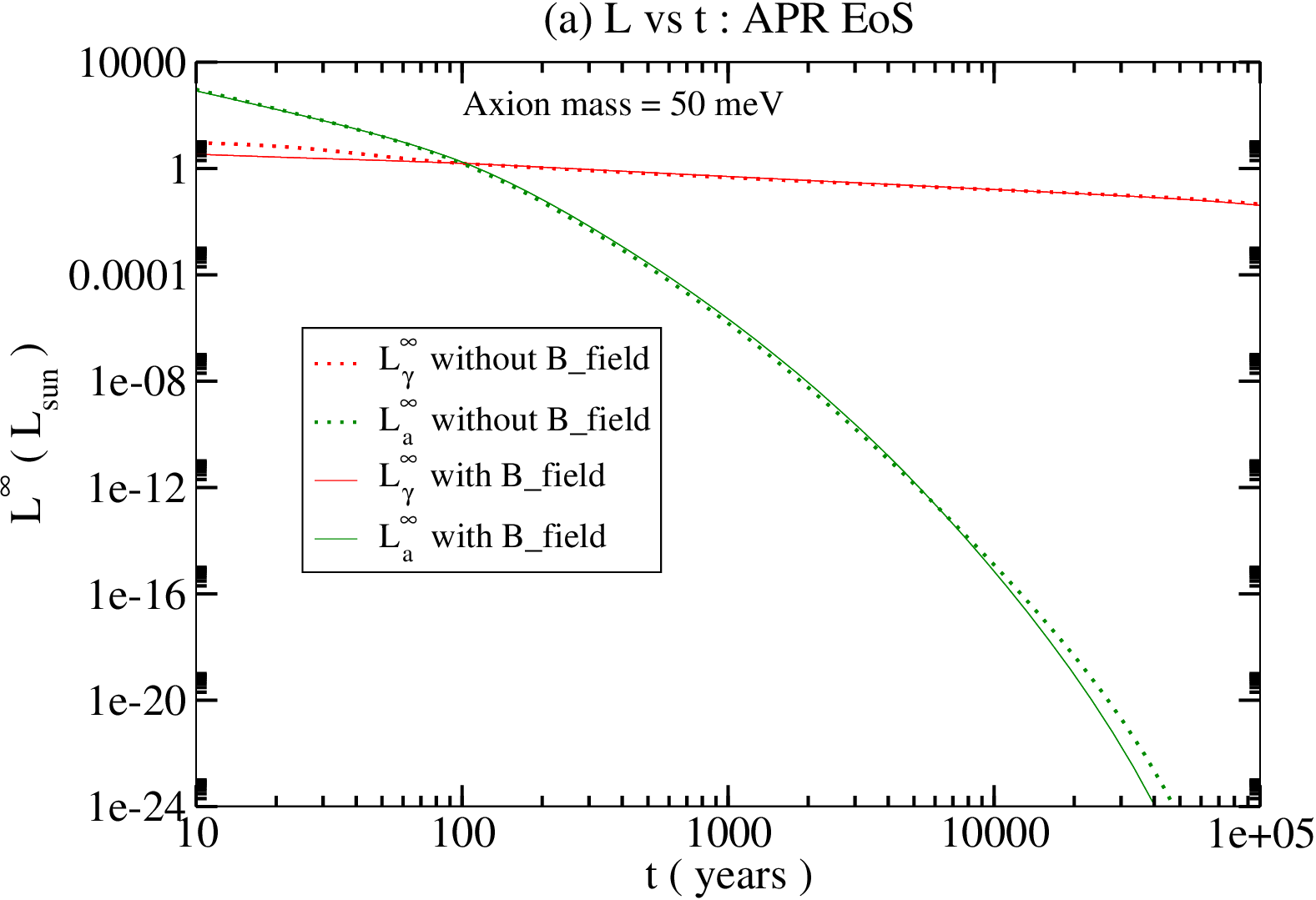}
\caption{The luminosity variation with time of axion mass $50$ meV for APR EoS in the presence and absence of magnetic field.}   
\label{fig:ltimevar1}
\end{figure*}

Figure(\ref{fig:ltimevar1}) shows photons, axions and neutrinos luminosities as a function of time for $50$ meV axion mass using APR EoS. At early times, the photon luminosity curve shows a sizeable difference between the results obtained in the presence and absence of magnetic fields. In the case of axion luminosity, this difference is not too much significant during the early times. The above figure also predicts that the axion luminosity and direct photon luminosity vanishes at a time slightly beyond $10^5$ yrs. 
The axion luminosity dominates over the photon luminosity upto time ~$\sim$ $10^2$ yrs and thereafter photon lumniosity always remains higher over the rest of time scales.  
Although strong magnetic fields do not change the qualitative behavior, the magnitude of the luminosity of direct photons \& axions is quite sensitive to the axion mass. 
 
\begin{figure*}[htb!]
\centering
\includegraphics[width=7.75cm]{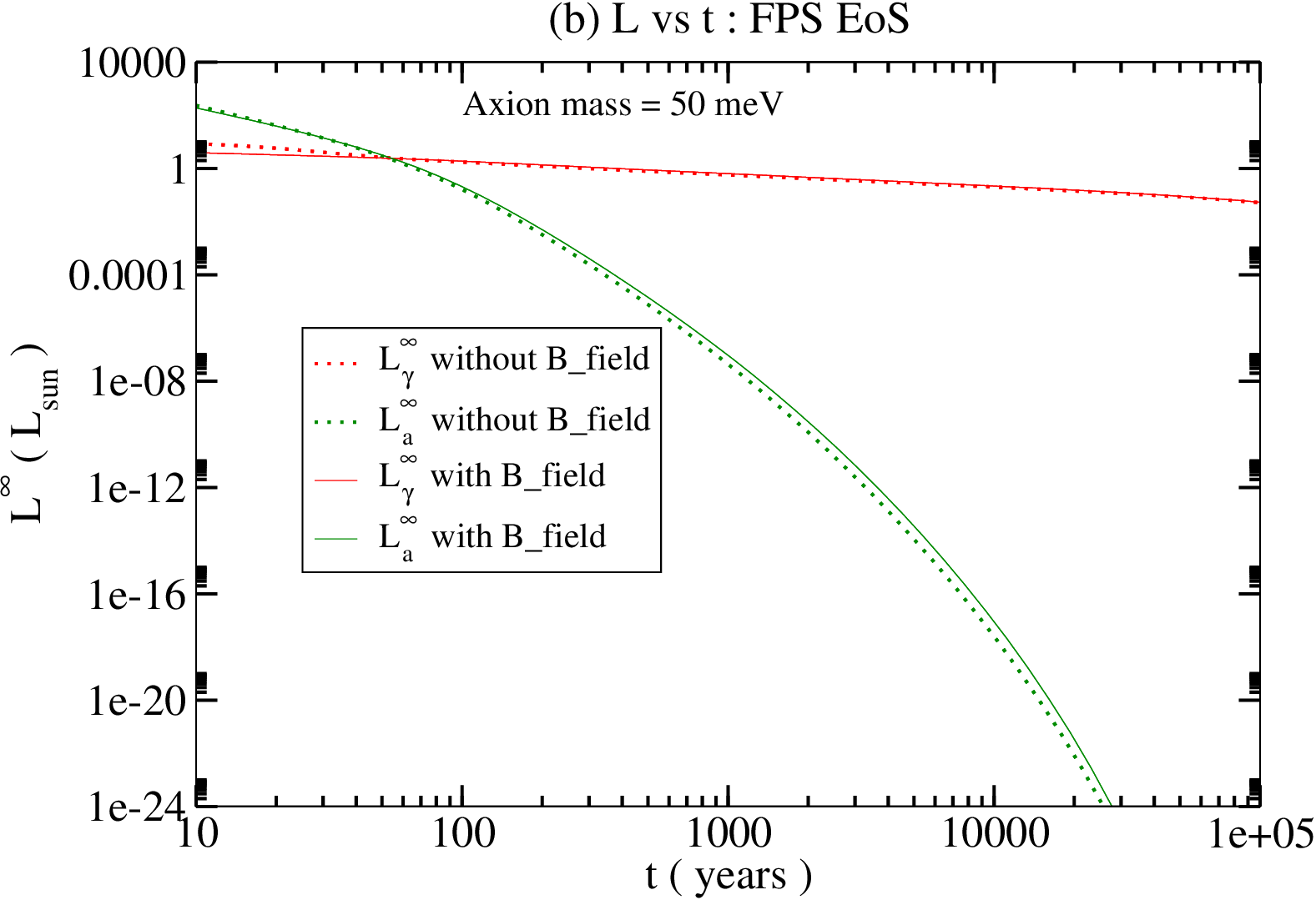}\ 
\caption{The luminosity variation with time of axion mass $50$ meV for FPS EoS in the presence and absence of magnetic field.} 
\label{fig:ltimevar2}
\end{figure*}

In Figure (\ref{fig:ltimevar2}), we have presented the luminosities versus time of the same axion mass $50$ meV for the FPS EoS. Here, also the qualitative behavior does not seem to change much due to the effect of the magnetic field. A significant departure can be seen in the photons and axions luminosities respectively in the presence of the magnetic field. Similar to the previous figure, during the early stages, the direct photon luminosity curve lies above the axion luminosity curve for times less than or equal to~$\sim$ $10^2$ yrs \\

\begin{figure*}[htp!]
\begin{tabular}{ccc}
\includegraphics[width=5.75cm]{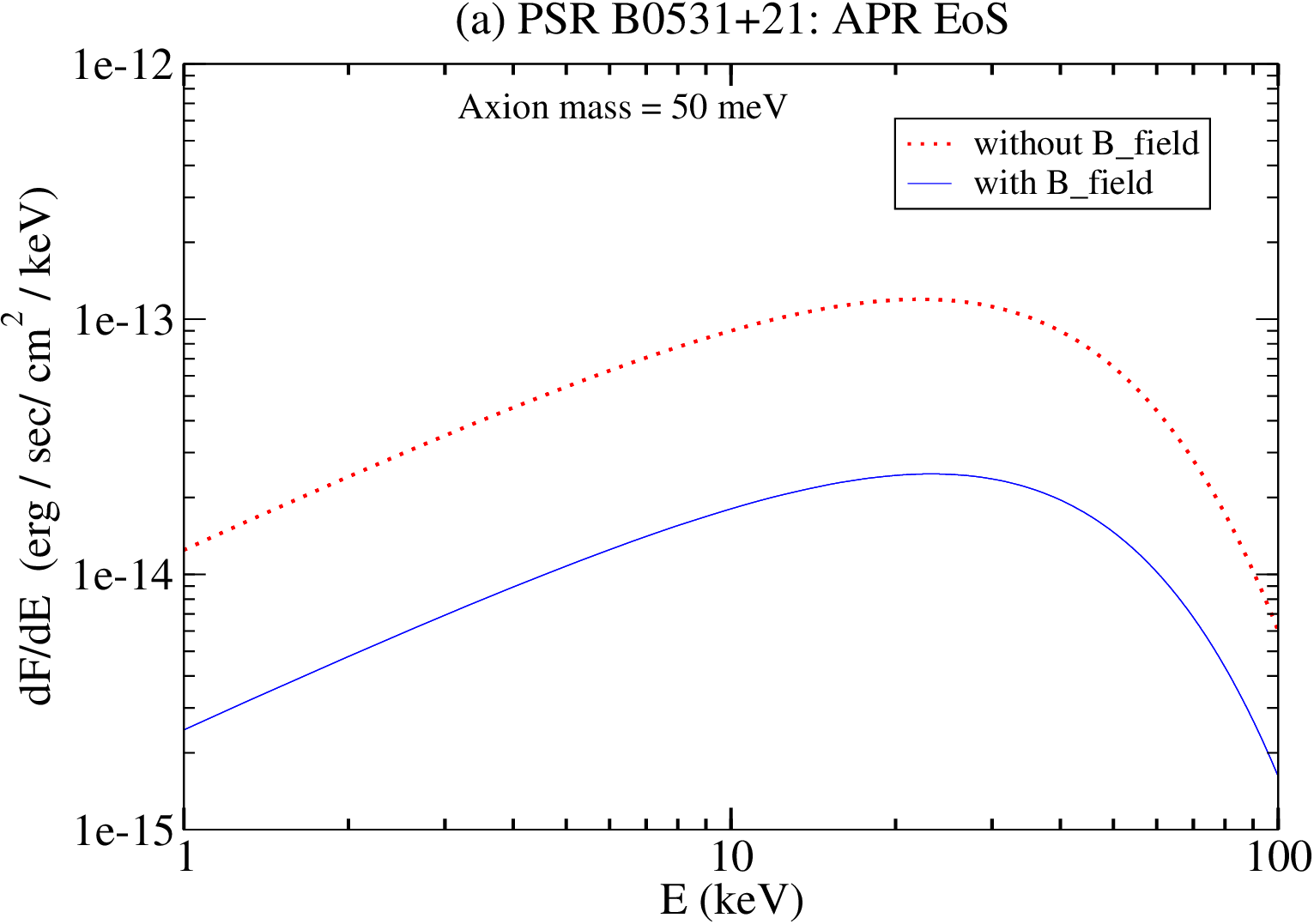} & \includegraphics[width=5.75cm]{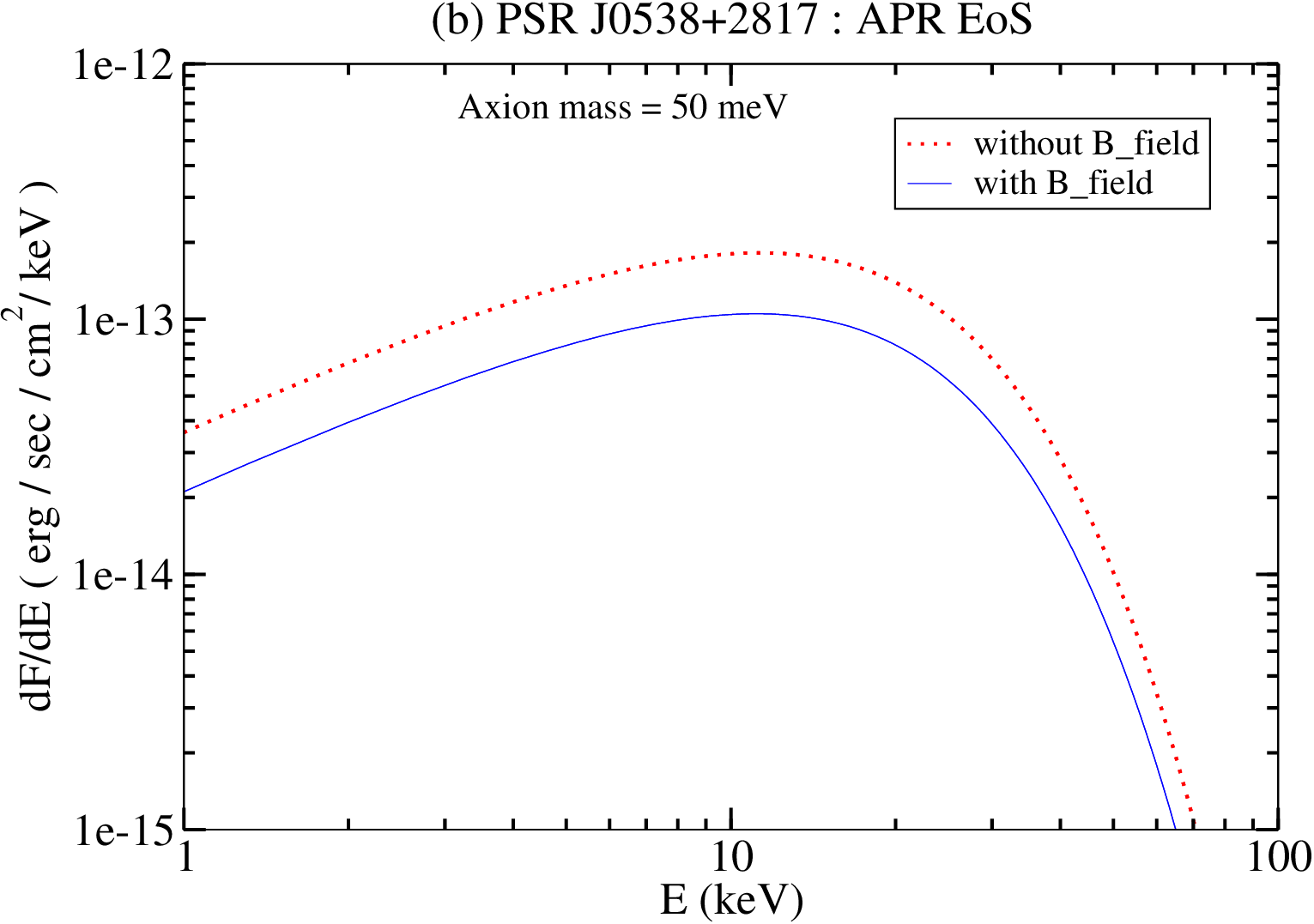} & \includegraphics[width=5.75cm]{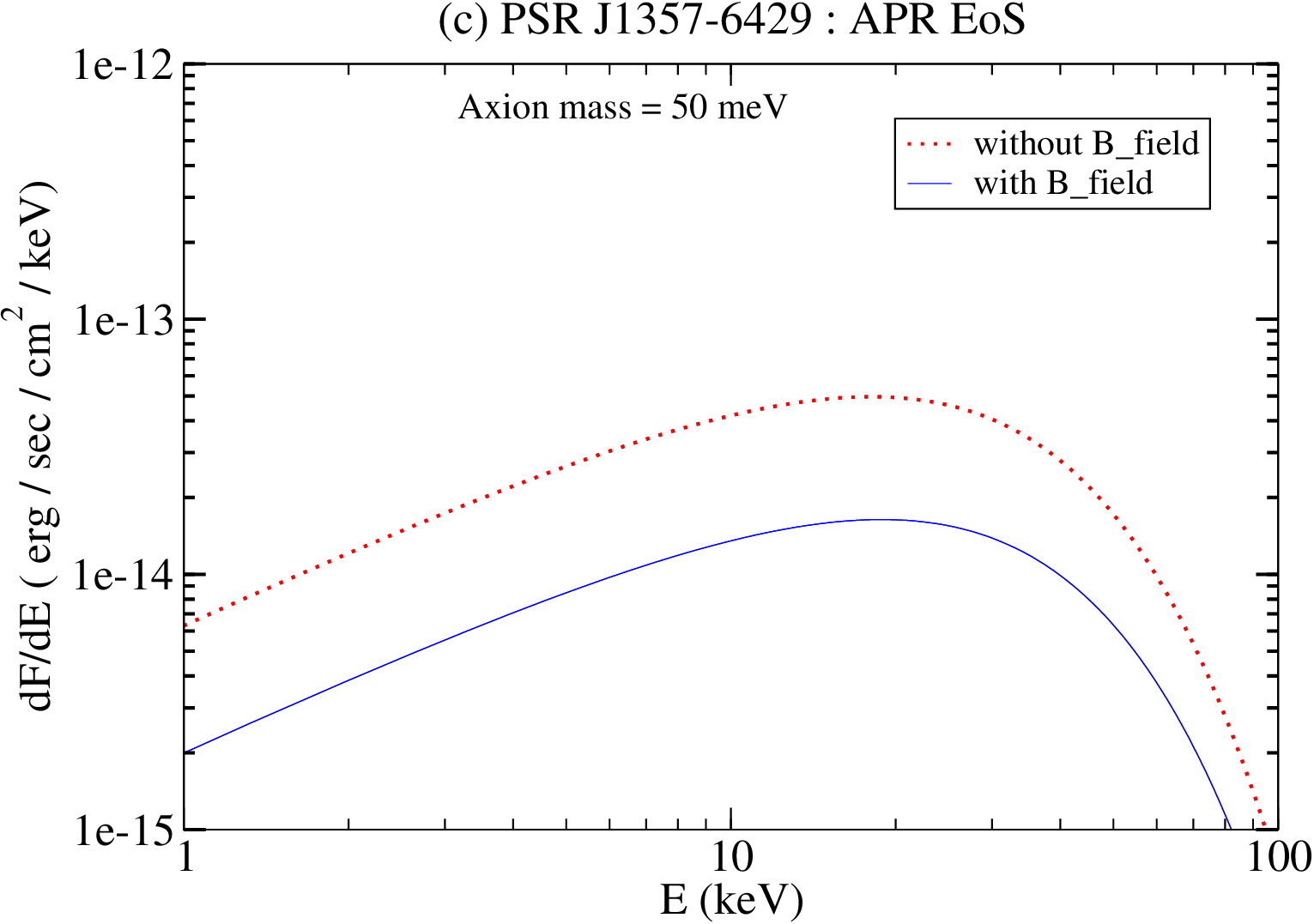} \\ \includegraphics[width=5.75cm]{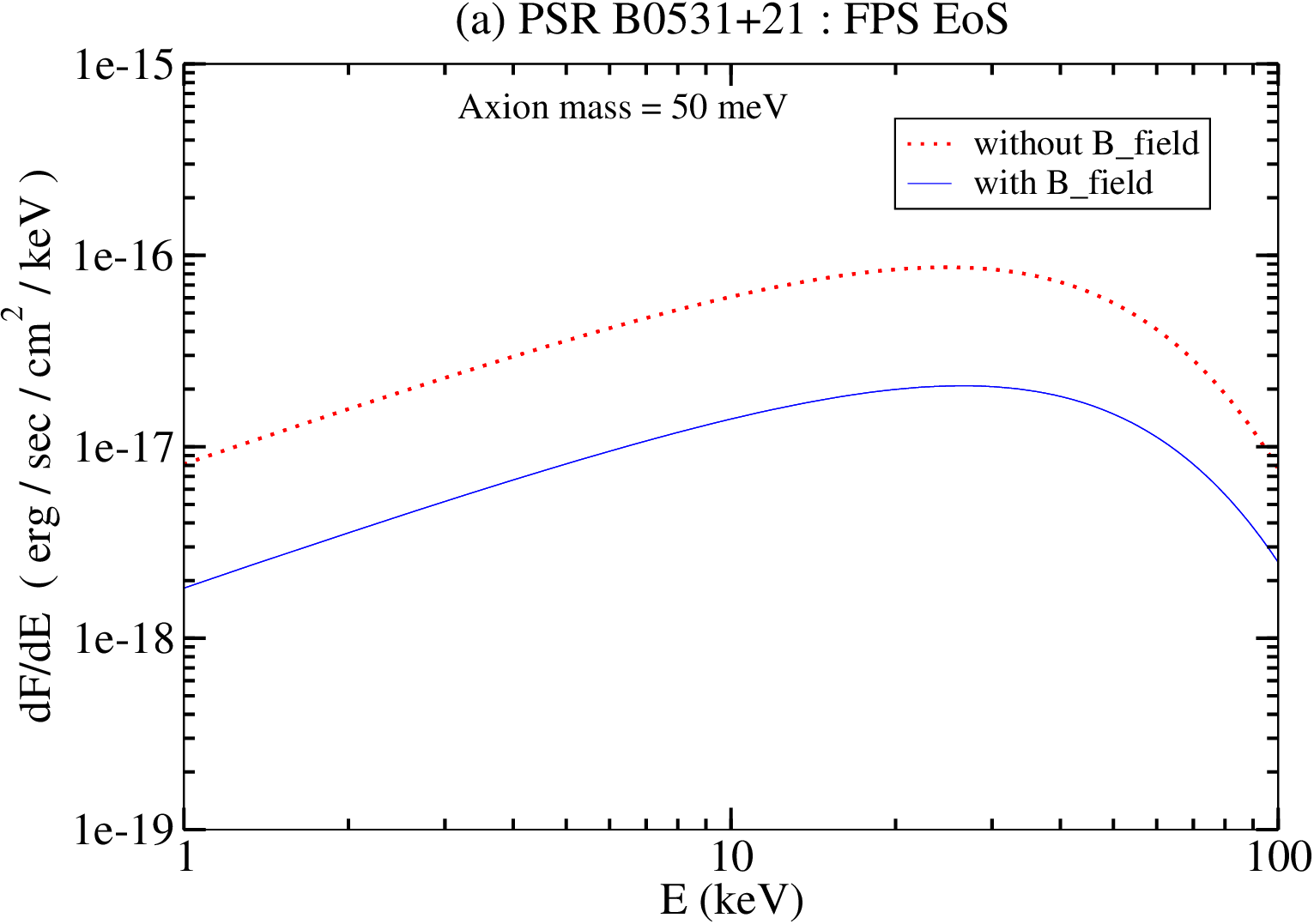} & \includegraphics[width=5.75cm]{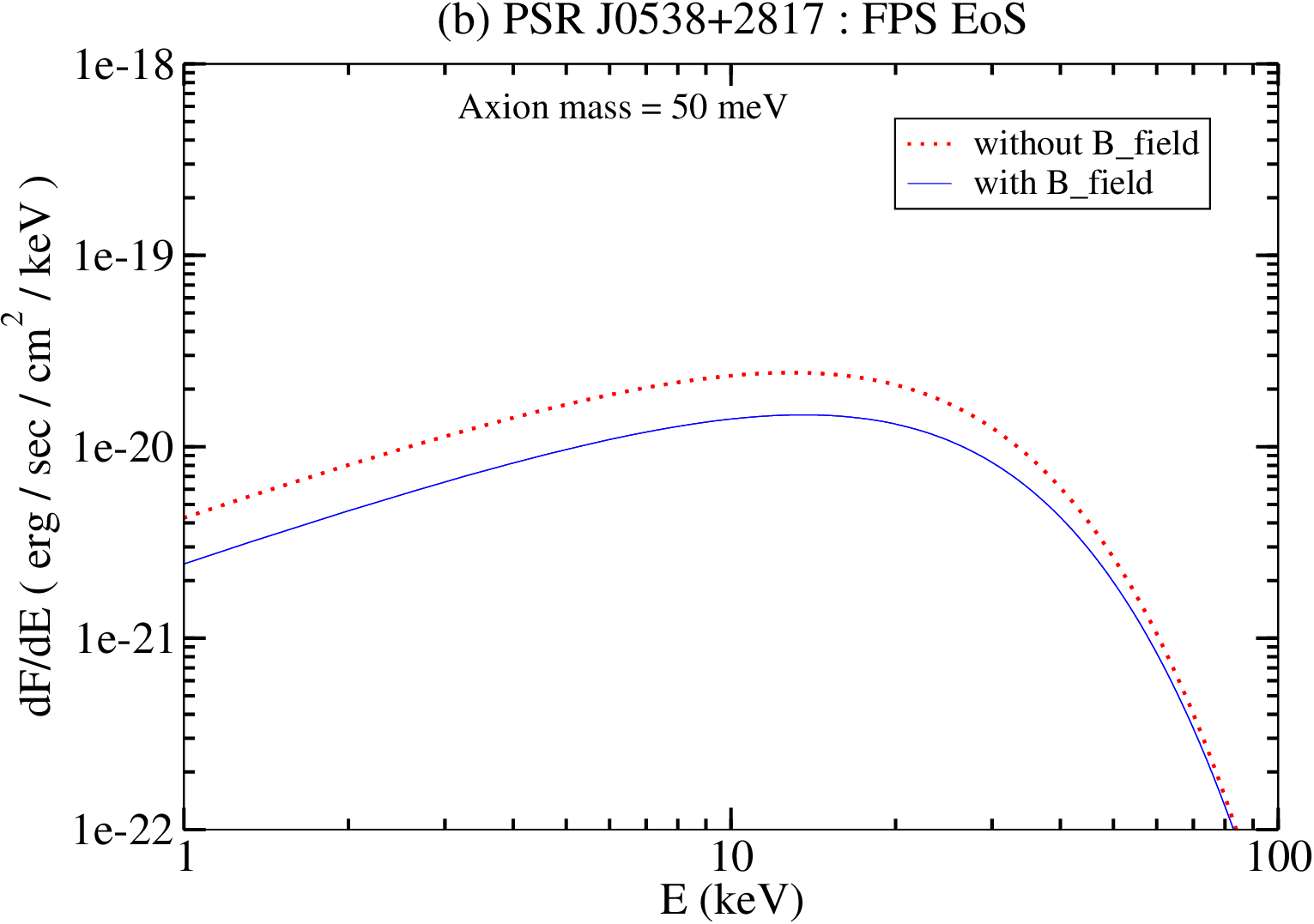}  & \includegraphics[width=5.75cm]{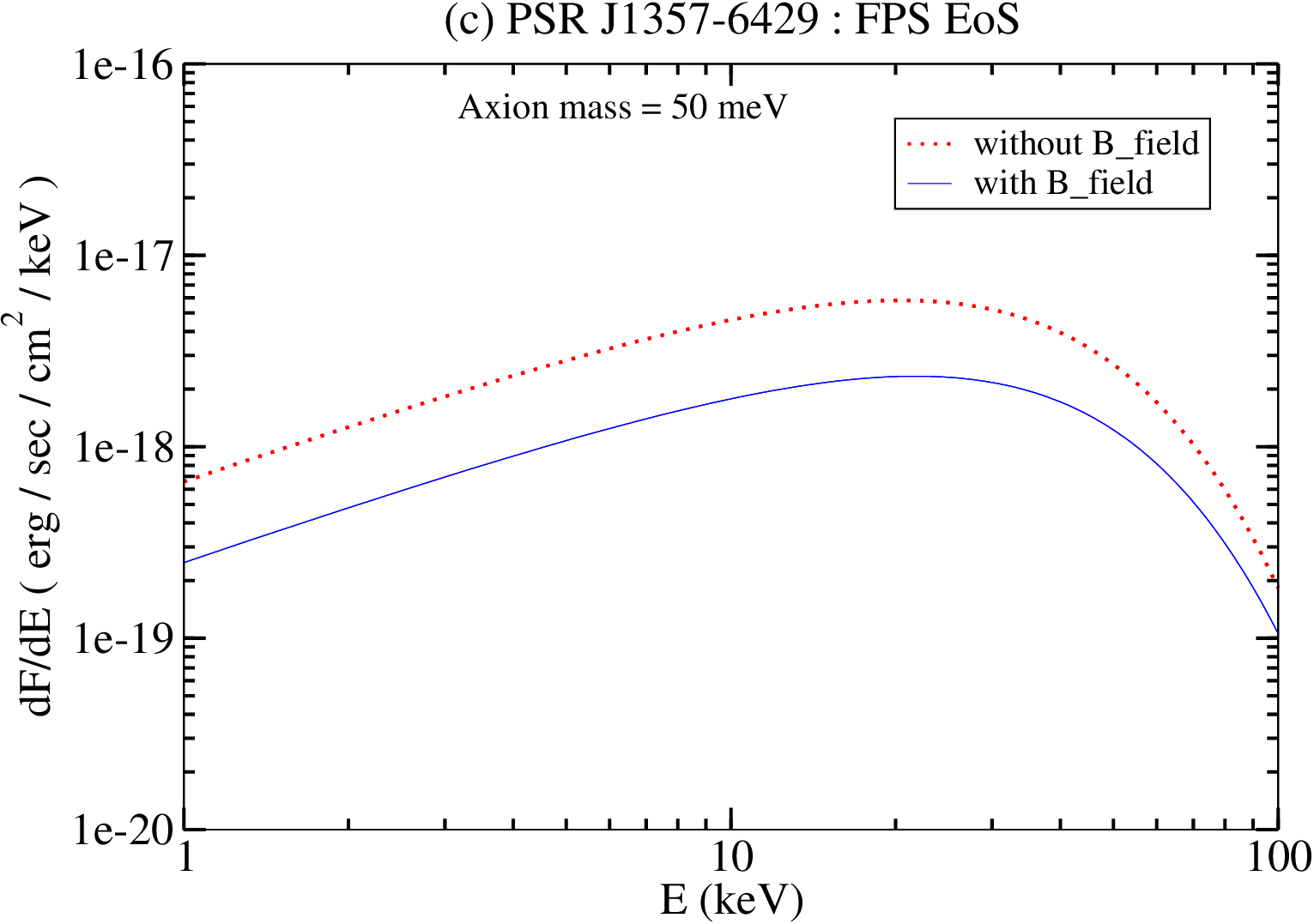}
\end{tabular}
\caption{The axion energy spectrum for three different NSs with characteristic ages $= 9.7\times 10^{2}$ yrs, 
$3.3 \times 10^4$ yrs and $7.3 \times 10^3$ yrs, respectively of axion mass $50$ meV in the presence and absence of magnetic field. The upper panel corresponds to APR EoS and the lower panel corresponds to FPS EoS}
\label{fig:nsradial1}  
\end{figure*}

The decrease in luminosities for direct photons, \& axions means the NSs core emits energy in lesser amount. This could happen due to various possible reasons, such as a decrease in temperature or changes in the internal processes occuring inside the NSs. 
The luminositites are all expressed in units of solar luminosity L$\odot$ = $3.826 \times 10^{33}$ ergs/s.
The photon luminosity at very short times are not very accurate as our time unit is coarse and
extrapolated.
In Figure (\ref{fig:nsradial1}), we have shown axion energy spectrum of axion mass $\sim 50$ meV  $\omega$ for three different NSs with characteristic ages $= 9.7\times 10^{2}$ yrs ~\cite{O'Dea_2014, refId0}, 
$3.3 \times 10^4$ yrs~\cite{Xu_2021,Ng_2007} and $7.3 \times 10^3$ yrs~\cite{Zavlin_2007,2012ASPC..466...29K}, respectively in the presence and absence of magnetic field. We have considered two different EoSs namely APR \& FPS EoS for the NS matter composition.

From the Figure (\ref{fig:nsradial1}), it is clear that in the presence of a magnetic field, the energy spectrum of axions undergoes a considerable changes. Without the magnetic field, the corresponding curve remains higher than a magnetic field inside the NSs. This is due to the decrease of luminosity of axions in the presence of magnetic field. For the APR EoS (upper panel) the axion energy spectrum peaks between $50-70$ keV axion energy and thereafter it starts decreasing with increase in the axion energy. The gap between curves corresponding to with and without magnetic field becomes narrower at higher axion energy.  
Although the APR and FPS EoS axion energy spectrum shows no qualitative different behavior. It is clear from the figures of FPS EoS (lower panel) that the difference between the energy spectrum from no magnetic case and with magnetic field case is small as compared to the case of APR EoS. Other behaviors concerning axion energy spectrum looks almost similar to the case of APR EoS.   

\begin{figure*}[htb!]
\begin{tabular}{ccc}
\includegraphics[width=5.75cm]{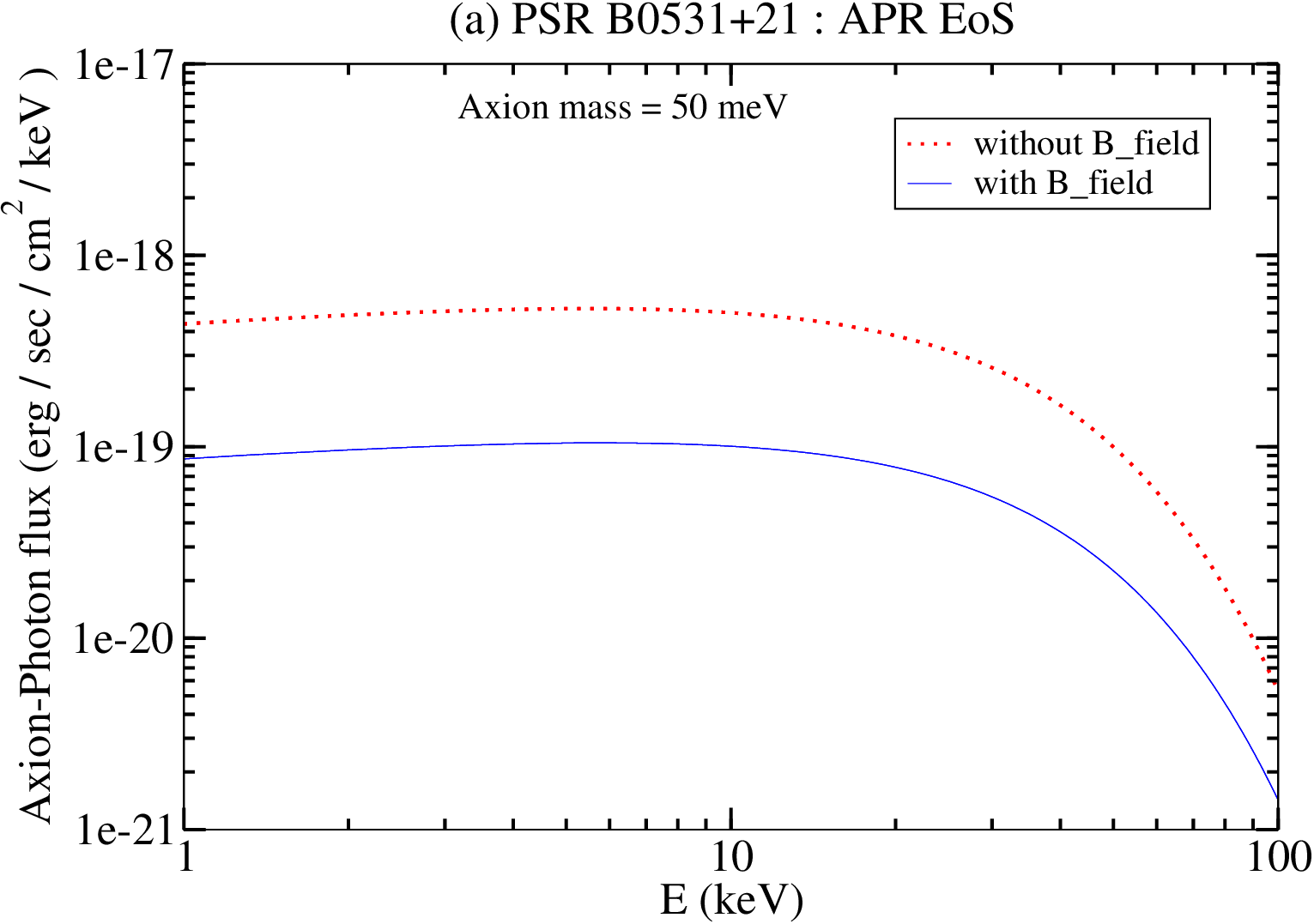} & \includegraphics[width=5.75cm]{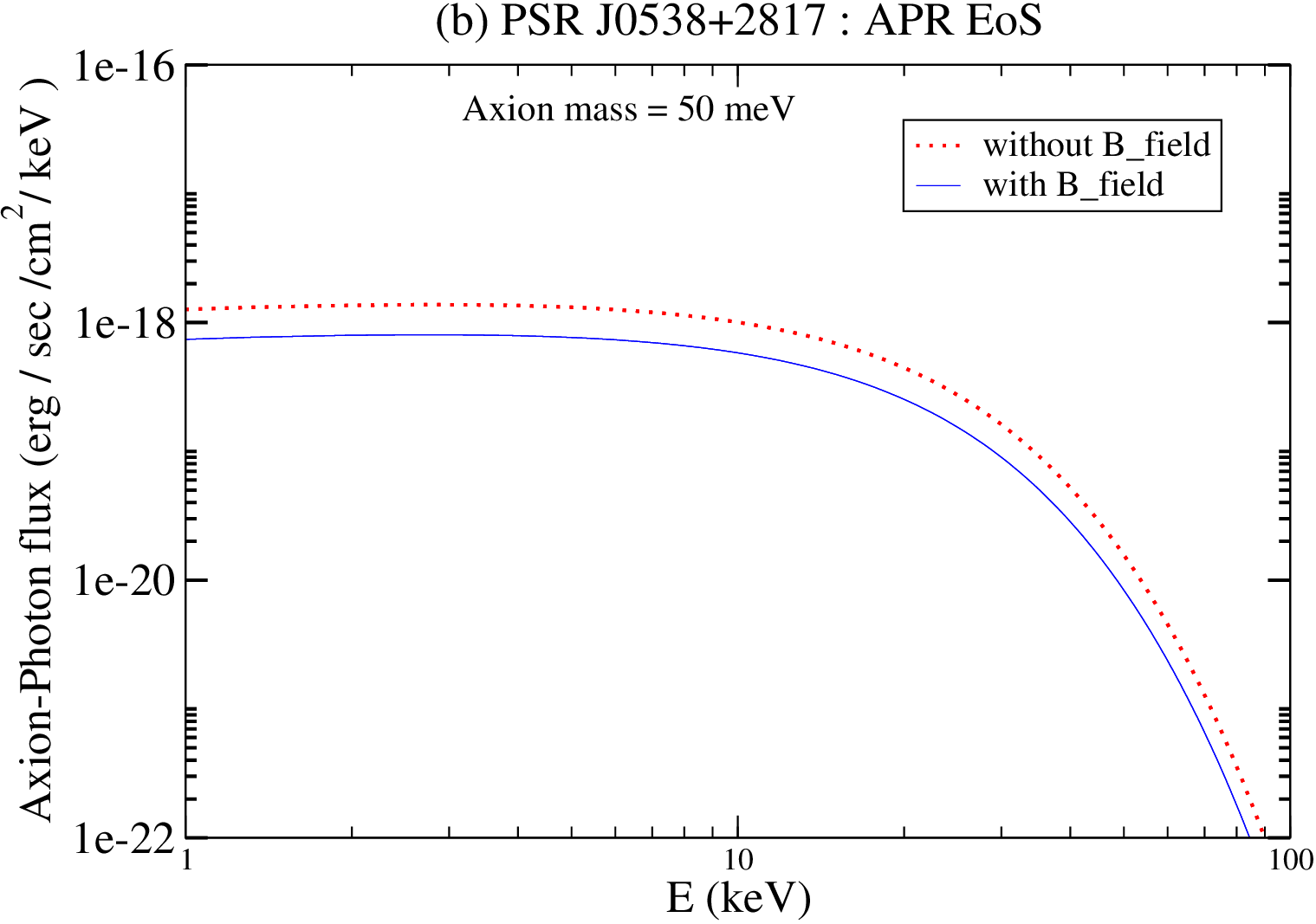} & \includegraphics[width=5.75cm]{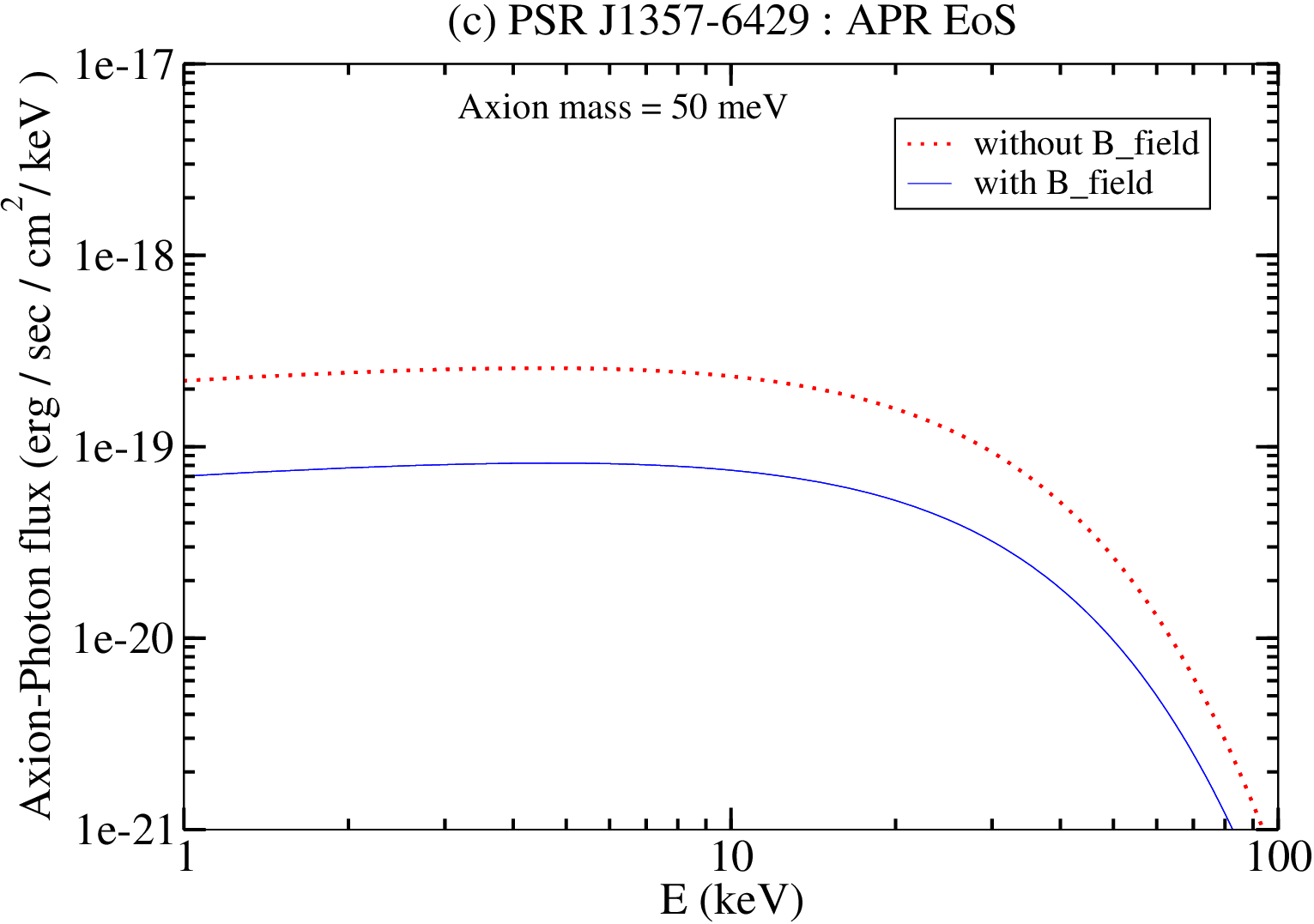} \\
\includegraphics[width=5.75cm]{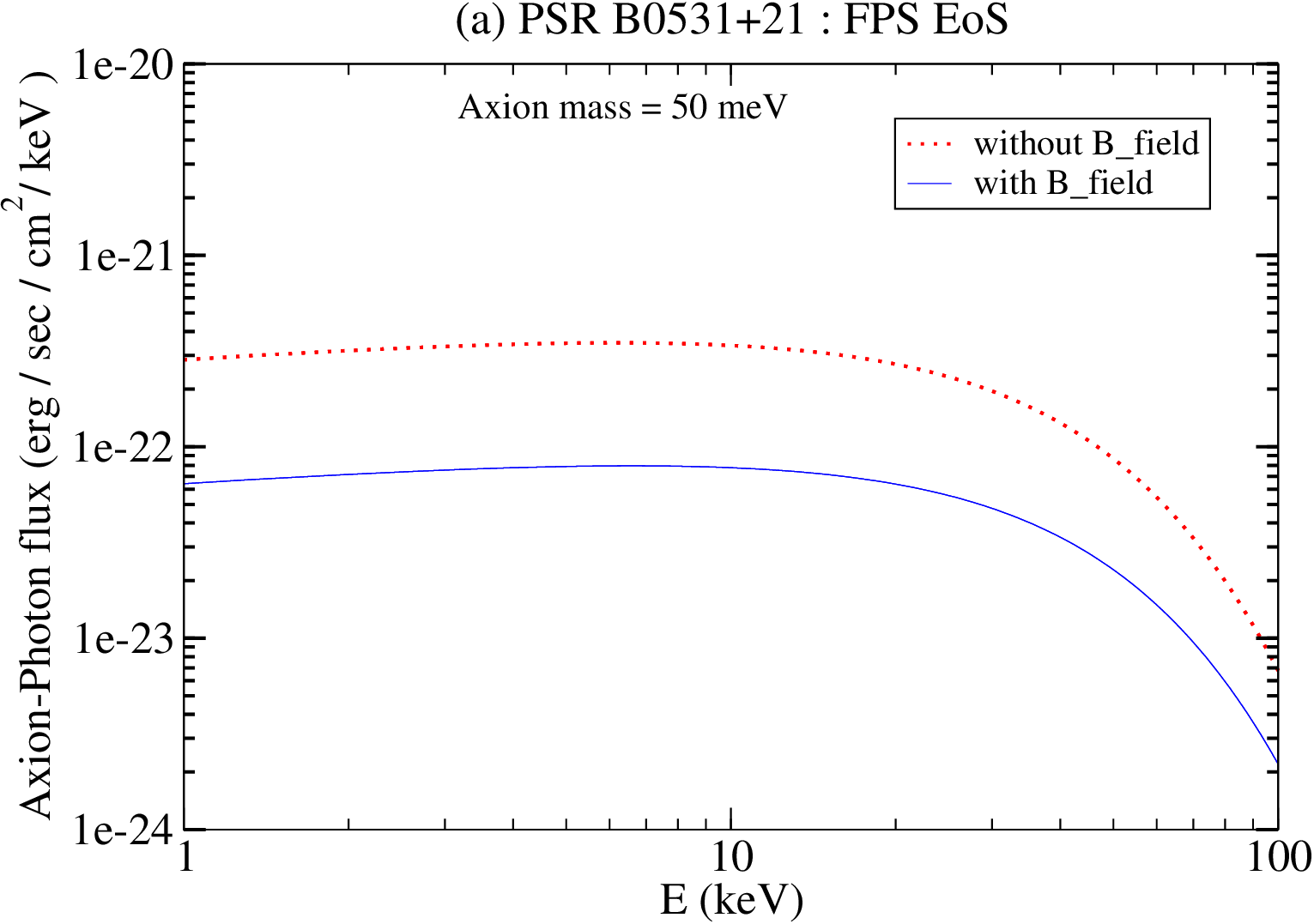} & \includegraphics[width=5.75cm]{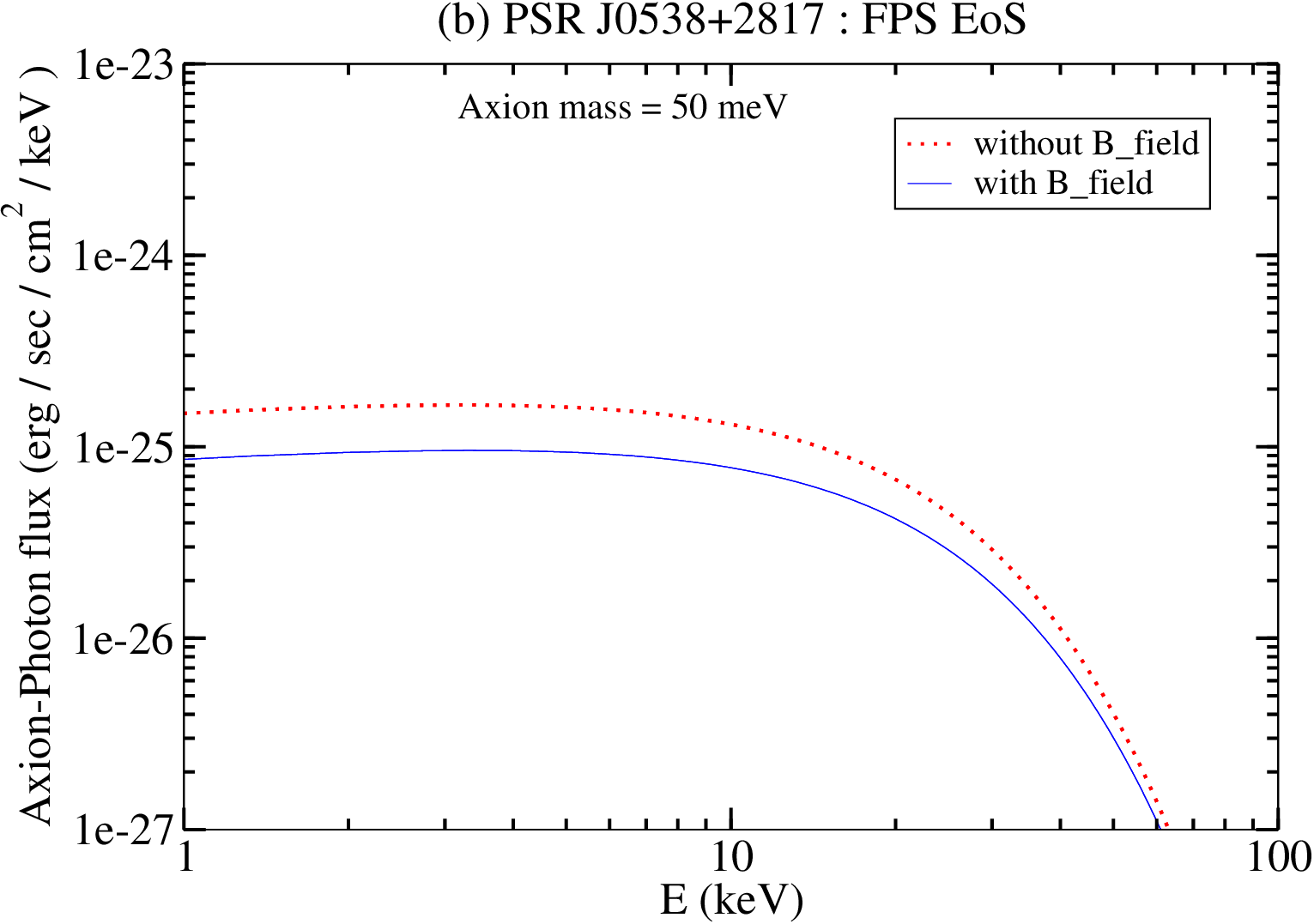}& \includegraphics[width=5.75cm]{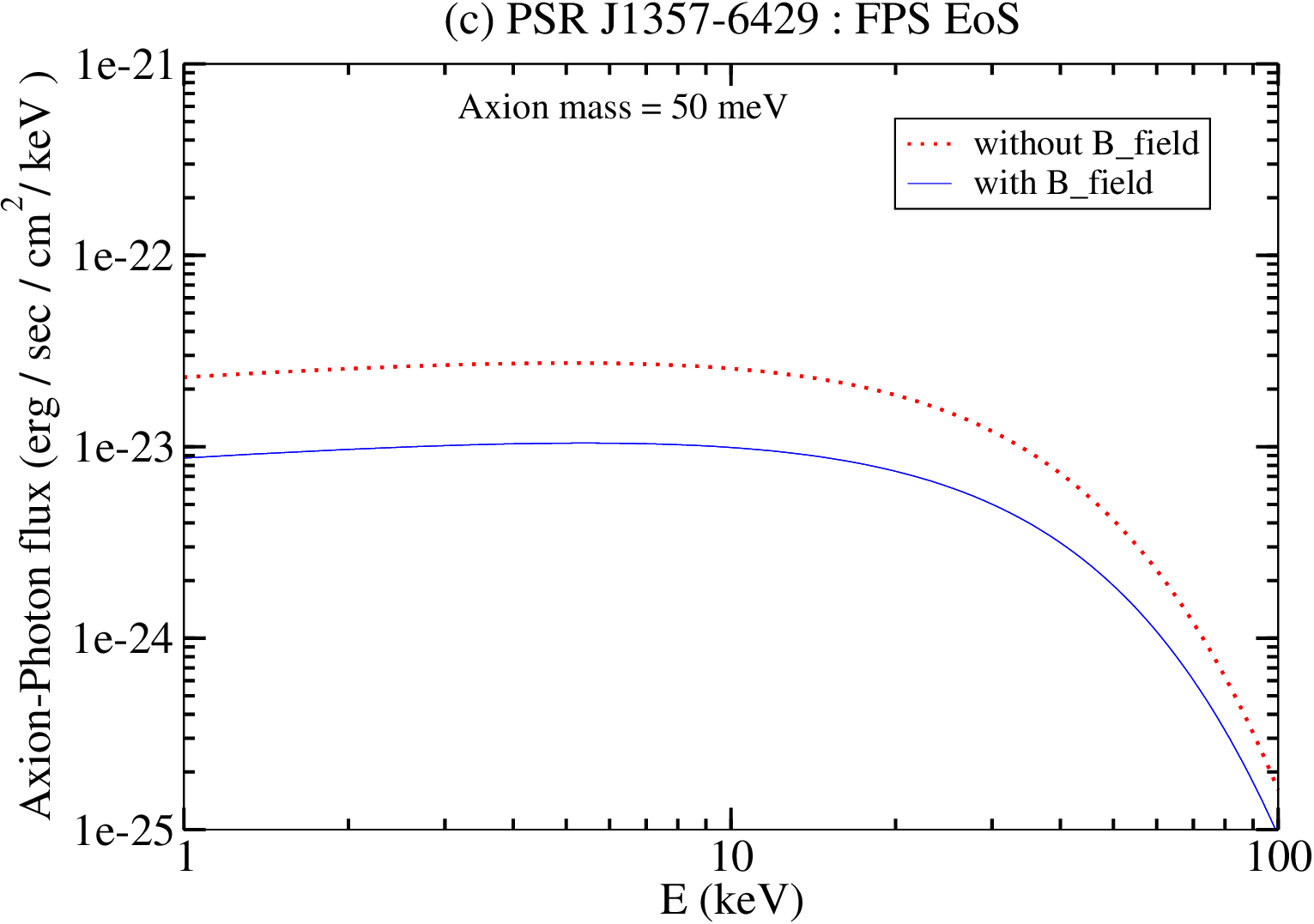}
\end{tabular}
\caption{The axions converted photons flux variation as a function of axion energy for three different NSs with Age $= 9.7\times 10^{2}$ yrs, $3.3 \times 10^4$ yrs and $7.3 \times 10^3$ yrs, respectively of axion mass $50$ meV in the presence and absence of magnetic field. The upper panel corresponds to APR EoS and the lower panel corresponds to FPS EoS} 
\label{fig:ltimevar4}
\end{figure*}

The upper panel of the figure (\ref{fig:ltimevar4}) depicts the variation of axions converted photons flux as function of the axion energy for three different NSs at axion mass $\sim 50$ meV for APR EoS. The figure shows a flat behavior for almost $30$ keV energy and then begins to decrease with an increase in axion energy. However, the axion converted photon flux with a magnetic field always remains lower than in the case of without magnetic field. It can also be concluded that the direct photon emission during the cooling of the NSs remains dominant as compared to the axions converted photons under the effect of the strong magnetic field.

The lower panel of the figure (\ref{fig:ltimevar4}) shows the variation of axions-converted photons flux as a function of the axion energy for FPS EoSs at axion mass $\sim 50$ meV with and without including the magnetic field. Here, we have also observed a significant difference due to the presence of the magnetic field. The departure from the case with the magnetic field decreases with the increase in the axion energy. After incorporating the magnetic field, the axions-converted photons curve lies below the corresponding curve for without magnetic field in the case of FPS EoSs. We emphasize here that axions cooling calculations are very approximate. We
consider only the mechanism of the singlet, $^1S_0$ PBF process which occurs mainly in the
core of an NSs. 

\section{Conclusions}
\label{conc}
In the present work, we have investigated the luminosities variations of strongly magnetized NSs as a function of time. Our investigations are based on the assumption that axions,neutrinos and photons are emitted from the core of NSs~\cite{Buschmann:2019pfp}. We have incorporated Cooper-pair-breaking formation process (PBF) for axions \& neutrinos production mechanism in the core of NSs in the presence \& absence of strong magnetic fields using the NSCool code. We have also explored the variation of the axion energy spectrum for a specific value of axion mass $50$ meV. Finally, we have determined the axions-to-photons conversion probability to obtain the emitted axion-converted photon flux for both without and with the magnetic field. It is an important observable and can be employed to constrain the various properties of axions. It could also be possible that these particles might shed light on its detection possibilities using astrophysical probes. The detection of axion dark matter could be possible through a narrow radio spectrum given out by neutron stars (NSs). Our comparison results confirm that an intense magnetic field significantly alters the axion's energy spectrum and axion-converted photon flux.  
 \vskip 1cm
\section*{Acknowledgments}
We are grateful to Dany Page, Malte Buschmann, and T. Opferkuch for clearing our doubts related to usage of NSCool code. SY thanks Birla Institute of Technology and Science, Pilani, Pilani Campus, Rajasthan, for financial support.

\bibliographystyle{apj}
\bibliography{references} 

\label{lastpage}

\end{document}